# Network structure of phonographic market with characteristic similarities between musicians

A. BUDA[a], A. JARYNOWSKI[b,c*],

[a]Wydawnictwo Niezależne,

67-200 Głogów, Oriona 15/8, Poland

[b]Smoluchowski Institute of Physics, Jagiellonian University,

30-059 Kraków, ul. Reymonta 4, Poland

[c]Department of Sociology, Stockholm University,

SE-106 91, Stockholm, Universitetsvägen 10 B: plan 9, Sweden

We investigate relations between best selling artists in last decade on phonographic market and from perspective of listeners by using the Social Network Analyzes. Starting network is obtained from the matrix of correlations between the world's best selling artists by considering the synchronous time evolution of weekly record sales. This method reveals the structure of phonographic market, but we claim that it has no impact on people who see relationship between artists and music genres. We compare 'sale' (based on correlation of record sales) or 'popularity' (based on data mining of the record charts) networks with 'similarity' (obtained mainly from survey within music experts opinion) and find no significant relations. We postulate that non-laminar phenomena on this specific market introduce turbulence to how people view relations of artists.

PACS numbers: 89.65.Gh, 05.45.–a, 42.65.Sf, 42.55.Px

1. Motivation and data sources

### 1a) History of phonography

People have played and listened to the music since ages but the origins of phonography came from Thomas Edison who developed many devices including the phonograph (1877). The gramophone record was one of the dominant audio recording formats throughout much of the 20th Century. The Italian tenor Enrico

---

* corresponding author; email: andrzej.jarynowski@uj.edu.pl

Caruso (1906) and The Beatles (1964) were the first who sold more than 1,000,000 copies of their records. The new formats (vinyl, cassette, compact disc and mp3, etc.) were more and more common. Thus, Michael Jackson's 'Thriller' has become the most popular record ever and sold over 110,000,000 copies [1].

**1b) Establishment of phonographic market**

Phonographic markets are well defined complex systems almost as old as financial markets. The figures speak for themselves: while Americans spent 100 million dollars on records before the fatal 1929 stock market crash, this number had plunged to a merely 6 million by 1933 [2]. Nowadays, global music sales in 2009 fell by 7% to US $17 billion. This is disappointing, but there are some very positive points. Digital music goes global in 2011 while action on piracy gains momentum. With rapid expansion into new markets by services such as iTunes, Spotify and Deezer, the major international digital music services are now present in 58 countries, compared to only 23 at the start of 2011 [3]. Traditionally, the record charts are based on weekly record sales. According to the IFPI (International Federation of the Phonographic Industry), the world's largest phonographic markets are [1,4]: the USA, Japan, Great Britain, France and Germany (see Appendix Fig. 0a). 80% of weekly record sales belong to the four biggest record companies (Universal, EMI, Sony BMG and Warner Bros [see Appendix Fig. 0B]). All the world's most popular artists are signed to these companies [1]. Thus, since 2003 it is possible to find their weekly record sales exactly [4]. Hence, the global phonographic market can be considered as a complex system and quantitative properties of phonographic markets have been already studied by economists and sociologists [5-8], and recently also by sociophysicists [9, 10]. In our research, we have chosen the portfolio of the world's most popular 30 artists (that sold more than 11 000 0000 units) according to albums sales between September 2003 and September 2011. These artists were popular, but not always critically acclaimed. Those artists represent variety of styles, genders,

and are signed to major record companies – see Appendix Fig. 0b.

### 1c) Role of experts in industry

Music writers also started treating pop and rock music seriously in 1964 after the breakthrough of the Beatles. Music press were more and more influential in discovering new artists. At the end of each year music magazines publish their critics polls that include the best albums and singles of the year according to critics opinions. These critics polls were often subjective, but had impact on weekly record sales [12]. We decided to study critics polls from 2003 to 2007. We chose the most influential music magazines (Mojo, NME, Pitchfork Media, Q, Rolling Stone, Spin, Uncut, Village Voice) and the Mercury Prize - the annual music prize awarded for the best album from the UK and Ireland. Nominations are chosen by a panel of musicians, music executives and journalists.  Music recommendation authorities seem to have influence on people's vision of music [11], so we decided to add it to our studies.

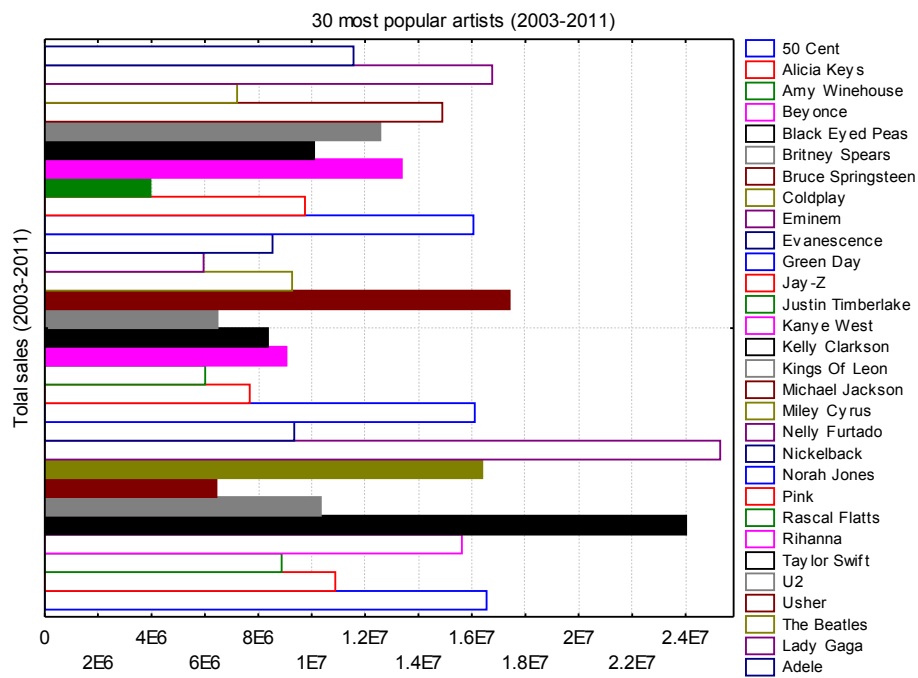

**Fig. 1)** Chosen cohort. The 30 best selling artists and music groups. The bands total record sales also contain their members contributions, e.g. Paul McCartney and John Lennon

record sales are included in The Beatles category.

### 1d) Research questions

In this chosen cohort displayed in Fig. 1, artists often collaborate in the same projects, or made the cover versions of other artists songs. A questions appear: how do similarity in music and cooperation between artists impact on connection in phonographic or popularity networks? What does influence on people's choices and decisions in buying records? Do people buy records like common good or like a luxury good? Do similar artist compete or are 'surfing' together on trends waves? How critics polls influence on phonographic market?

## 2. Introduction to analysis

The motivation of the present study is to find a kind of more general arrangement between artists and built networks and compare different approaches. The phonographic network is obtained starting from the matrix of correlations coefficient computed between the world's most popular 30 artists (this list correspond to all analyzes in this paper). The value of an artist is defined by weekly record sales. The graph is obtained starting from the matrix of correlations coefficient computed between the world's most popular 30 artists by considering the synchronous time evolution of the difference of the logarithm of weekly record sales. We show topological arrangement of market by hierarchical diagrams (Fig. 2) and networks (Figs. 5, 6). On the other hand, popularity of artist calculated from ranks in charts of their albums sales year by year was visualized on the hierarchical diagram in Fig. 3. Most influential music journal for different branches of music were investigated via data-minig procedures to find names or aliases of chosen artists with their rank on charts. The link between artists appears when both artists exist on the same chart list in a chosen year. An indicated relation is stronger if their rank on such a list is similar. Weights of the link grow if those artists appear in more than one chart that year. On the other hand, links between artists can be defined as a similarity between them. This similarity can be understood in two ways: as an automatically annotated links found on allmusic.com website (Fig. 7) or as an expert (subjective people's view [Fig. 8]). On allmusic.com there are two categories to link artists: following or followed by. That gives opportunity to obtain directed network. Experts, however, could rank their subjective feeling of similarity between artist in survey (Appendix Fig. 0c). They were asked at the end, what were for them the most important factors to link artists. The last step was to compare such obtained networks (Tables 2, 3).

## 3.     Hierarchical analyze of phonographic market and charts

Initially, we analyzed correlations network of artist's record sales. The correlation coefficient defines a degree of similarity between the synchronous time evolution of a pair of assets, where we took of underlying[1] sales value. We looked at correlation coefficient in whole period as well as in one years times windows, because of seasonality of artist fame and chance to compare it with others. Year window will also correspond to popularity measure, which is based on yearly charts.

A typical division of correlation, useful in finding the life-time of correlations between stocks in financial markets [13] where almost all correlation coefficients are positive, is not valid any more because of different properties of the system.

---

1  Initially, we worked on logarithmic returns as it is traditionally done in econophysics, but we decide to use underlying records sale to get better feeling of significance of obtained coefficients (significance in such a case indicates direct relationship and indirect in terms of returns. How exactly procedure of  building networks from correlation matrix can be found in Mantegna's book [13] or Buda's book [14].

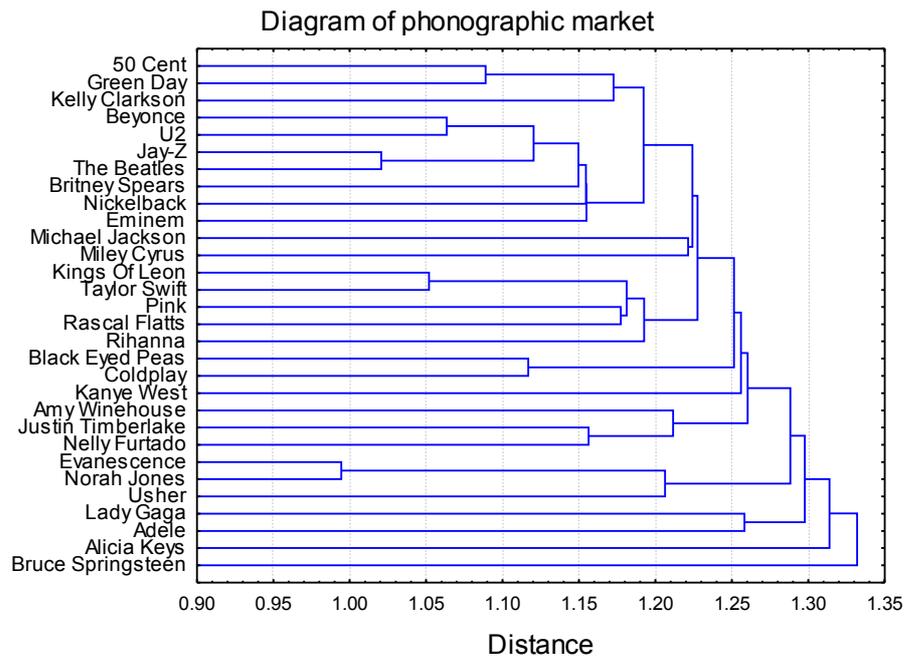

**Fig. 2)** Hierarchical structure of phonographic market (2003-2011).

Most of phonographic correlations are insignificant (85% of cases with $\alpha = 0.05$) but positive. However, some of the significant correlation coefficients in the phonographic markets are negative when artists essentially compete over the same group of customers.

Coincidence in presence on the same chart list in specific year was introduced as a measure of popularity. Accumulated coincidences matrix for 2003-2006 can be also presented as a hierarchical tree (Fig. 3) which seems to be different than sale tree (Fig. 2), but still reveals 'stars' cluster described more in chapter about SNA.

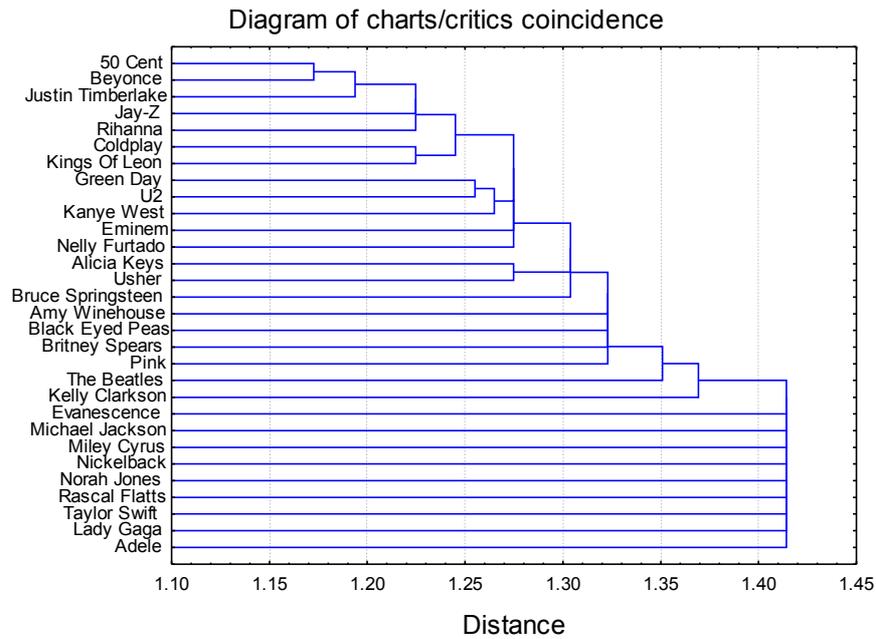

**Fig. 3)** Hierarchical structure of charts popularity (2003-2007).

Phonographic diagram can be also represented as a network. Correlation coefficient has then meaning of a weight of a link. Due to that definition, we would obey always full connected network. To avoid that, we decided to state some threshold. The best, natural one, in our opinion, for correlations is P-Value above certain α=0.05 and we are using the same rule everywhere in this paper when we are projecting a correlation matrix into a network.

**Tab. 1)** Correlations between binary networks obtained from record sales in next years from 2003 to 2011 with full period. Those significant (α = 0.05) are in bold.

| Corr. | full | 2004 | 2005 | 2006 | 2007 | 2008 | 2009 | 2010 | 2011 |
|---|---|---|---|---|---|---|---|---|---|
| full | 1.000 | -0.013 | **0.163** | **0.120** | **0.151** | -0.069 | 0.064 | -0.013 | 0.073 |
| 2004 | -0.013 | 1.000 | 0.029 | -0.035 | -0.039 | -0.024 | -0.056 | 0.010 | -0.032 |
| 2005 | **0.163** | 0.029 | 1.000 | 0.037 | -0.013 | -0.045 | -0.031 | 0.035 | -0.016 |
| 2006 | **0.120** | -0.035 | 0.037 | 1.000 | -0.019 | -0.003 | **0.110** | -0.048 | -0.007 |
| 2007 | **0.151** | -0.039 | -0.013 | -0.019 | 1.000 | 0.017 | -0.018 | 0.082 | -0.015 |
| 2008 | -0.069 | -0.024 | -0.045 | -0.003 | 0.017 | 1.000 | -0.034 | -0.004 | 0.035 |
| 2009 | 0.064 | -0.056 | -0.031 | **0.110** | -0.018 | -0.034 | 1.000 | **0.103** | **0.115** |
| 2010 | -0.013 | 0.010 | 0.035 | -0.048 | 0.082 | -0.004 | **0.103** | 1.000 | **0.101** |
| 2011 | 0.073 | -0.032 | -0.016 | -0.007 | -0.015 | 0.035 | **0.115** | **0.101** | 1.000 |

We checked how such a network evolve in time (Table 1). To do so, we divided time series in years interval (there are some economic reason why one year is the best interval and explanations comes from seasonal waves of selling, e.g. before Christmas or Valentine's Day average sales grow up rapidly). This network is not stable. If we look at correlations between binary networks (link exists if correlation between this two artist is significant in given time interval), there is only small memory, because only 29% of correlations between following years are significant (10% for all cases). Those binary network do not consider sign of correlation at all. That brings curiosity, what should be taken as a weight of the link. In standard analyze [12] correlation coefficient is transformed to Euclidean distance, and this transformation does not care if coefficient is positive or negative. To have in mind, that sign of correlation does matter, let us introduce two measurement: pure coefficient and its square. A square has another good property – it is underestimating small correlations, which plays a role of noise reductor.

4. **Product life-cycle in phonographic market**

The Black Eyed Piece record sales (Fig. 4) represent typical behavior of music label policy. Firstly, we should realize, what is happening in the date of a premiere of a new album of a famous artist. The rapid increase of record sales in such a week is obvious, but there is also a small increase just before the premiere. Fans are waiting for a new album and they are buying older ones. This is the stage of low growth rate of sales as the product in newly launched in the market. With a premiere of an album the first single of that album comes to promote it. After some time, next singles are coming and if it has got popularity the record sale increases. A growth comes with the acceptance of the innovation (new album) in the market and profit starts to flow. The last stage of product lifecycle is maturity. In this end stage of the growth rate, sales slowdown as the product have already achieved it acceptance in the market. Therefore, new artists start experimenting in order to compete by innovating new models of the product [15, 16]. Thus, The Black Eyed Peas should release another albums to be still on the top of the record charts. It is clearly shown in their record sales history.

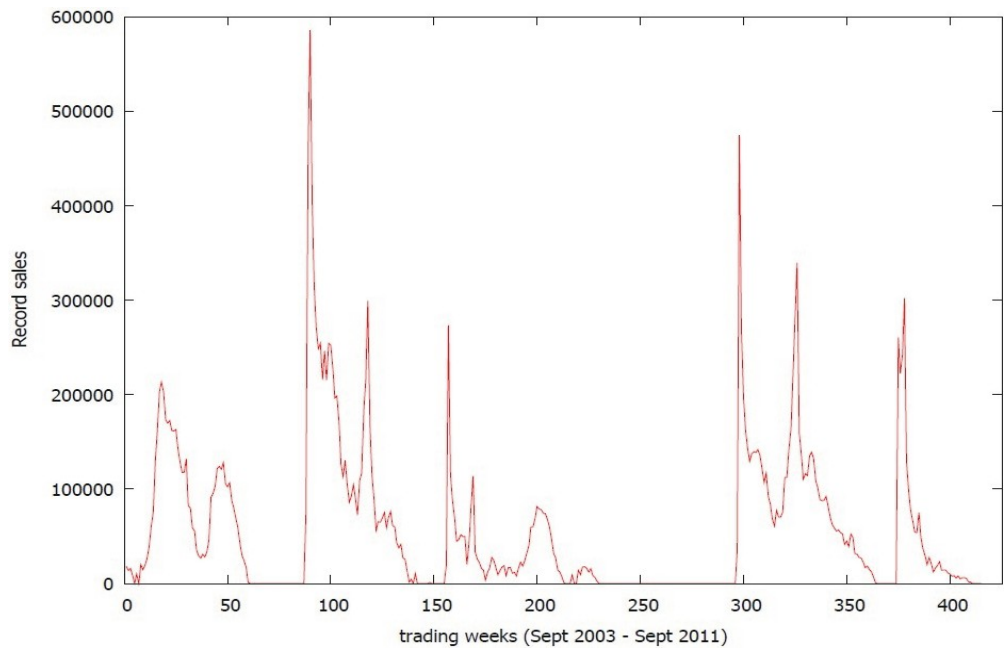

**Fig. 4)** The Black Eyed Peas record sales history.

5.   **Network analysis**

For that, we have built the network for portfolio of artists. It provides an arrangement of assets, which selects the most relevant connections of each point of the set. Although the network structure in financial markets reflects the classifications of stocks in the industry sectors and sub-sectors reported in the Forbes annual reports on American Industry [17], The Social Network Analysis of correlations between artists does not always fit to music genres classified by the Billboard and other music magazines devoted to music industry.

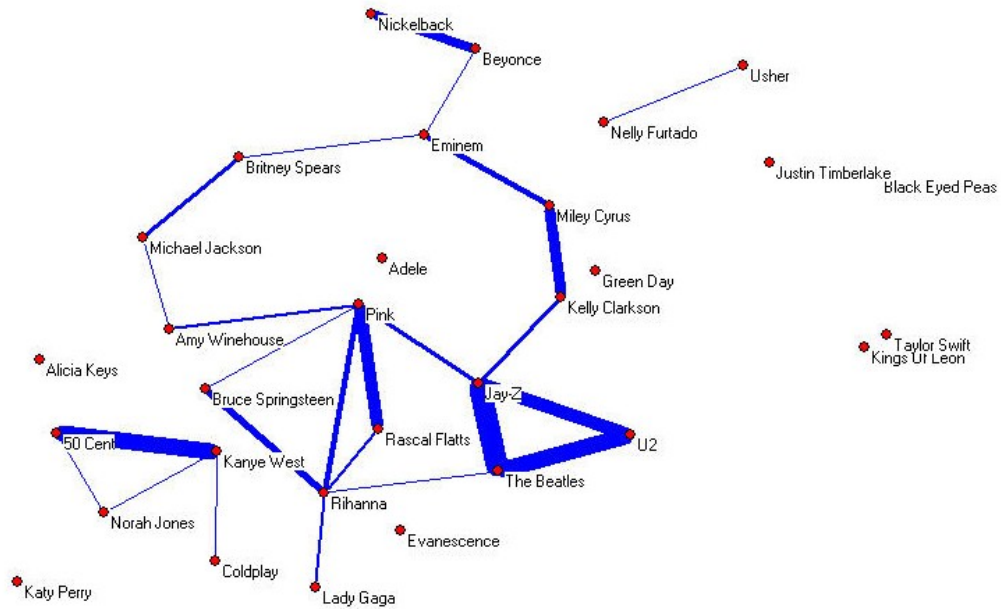

**Fig. 5)** Significant (α = 0.05) positive correlation between artist sales.

The SNA reveals [Fig. 5] sectors that belong to rap (Kanye West, 50 Cent), but does not show the sector for pure pop music. Instead of pop, the main community, we have a celebrity sector that contain Lady Gaga, Rihanna, Bruce Springsteen, Pink and superstars Jay-Z, The Beatles and U2. What do these artists have in common? Although they represent various styles and genres, the only common thing they have is fame, high record sales, popularity and a place in music history.

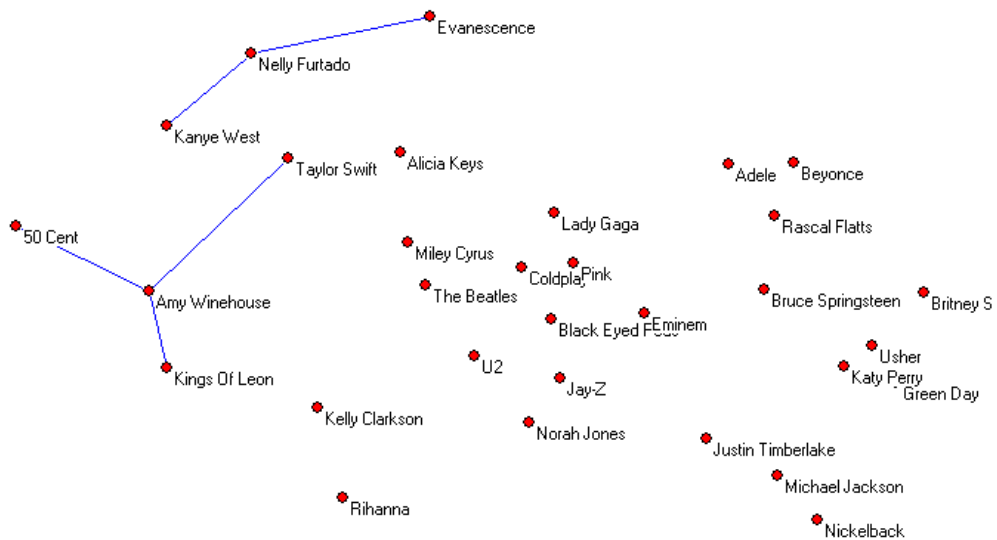

**Fig. 6)** Negative correlations (α = 0.05) between artists sales.

Negative correlations network is even more difficult to analyze. There is Amy Whitehouse, who had mainly negative correlations (Fig. 6). They could have come from her sudden death and kick her out from a normal system state.

## 6. Methods of comparison

We compared sales networks with such networks that explain similarity. Allmusic.com portal allows us to built a directed graph, because it provides two kinds of relations: following and followed by. All other networks are undirected so in comparison with them, we took accumulation of arrows as a weight of the links. This is standard action to project directed network into undirected one [18].

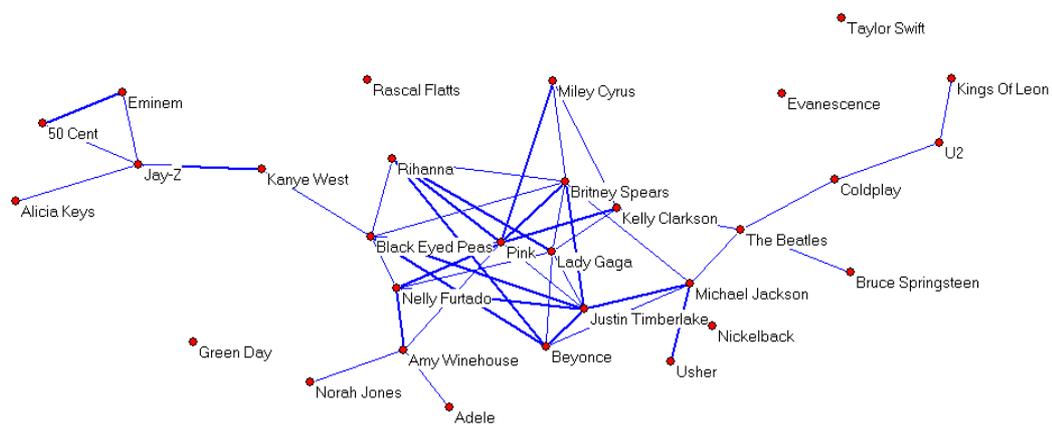

**Fig. 7)** Allmusic.com similarities.

Additional to that, we ask some music experts to annotate their ideas of links between artists. We used surveys to do so (Appendix Fig. 0c). To obtain expert view, we took an average network from every entry to survey[2] we were running. Experts were also asked how they rank relations. For example, A. Buda based on this network on his book about music history (Fig. 8).

---
2  The survey ( http://gulakov.cba.pl/artysci.php ) is still open for anybody who wish to participate in future projects.

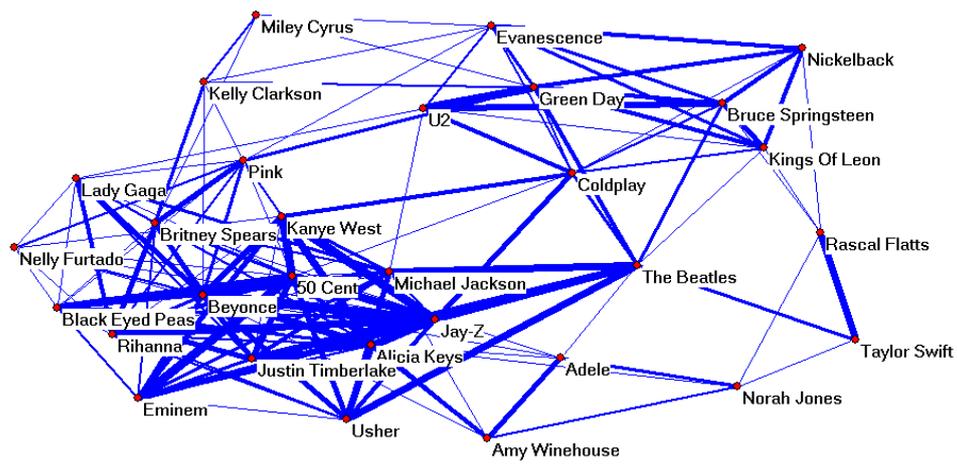

**Fig. 8)** Expert's (Andrzej Buda) view.

If we compare all of naturally weighted networks (Table 2), we could see that there is no relation between similarity networks and phonographic ones (does not matter whether networks are based on a pure correlation coefficient or its square).

**Tab. 2)** Correlation between networks 2003-2011.

| Networks | 2003-2011 | | | | |
| --- | --- | --- | --- | --- | --- |
| | expert (Buda) | experts | all_music | market_corr | market_sq |
| expert (Buda) | 1.00 | 0.79 | 0.34 | -0.01 | 0.02 |
| experts | 0.79 | 1.00 | 0.41 | -0.01 | -0.01 |
| all_music | 0.34 | 0.41 | 1.00 | 0.02 | 0.02 |
| market_corr | -0.01 | -0.01 | 0.02 | 1.00 | 0.93 |
| market_sq | 0.02 | -0.01 | 0.02 | 0.93 | 1.00 |

A lack of correlation between sales and similarity can be understood easily from the point of view of economics. On the other hand, a lack of a relation between sales and popularity (Table 3) was a surprise (both networks are establishing music market). Unfortunately, sale and critic (chart) networks do not cover the same time window and precision is different (sales record-every week, charts-every year). To imitate a year precision in the case of sales, we calculated correlations between artist's sales for every year and treat networks as a binary ones (if correlation is significant, then the link exists) and accumulate them over years (2003-2007), for which we ware able to check ranks in charts.

**Tab. 3)** Correlation between networks 2003-2006.

| Networks | 2003-2007 | | | |
|---|---|---|---|---|
| | experts | all_music | critics | market_cum |
| experts | 1.00 | 0.42 | -0.00 | -0.02 |
| all_music | 0.42 | 1.00 | -0.05 | 0.00 |
| critics | -0.00 | -0.05 | 1.00 | 0.01 |
| market_cum | -0.02 | 0.00 | 0.01 | 1.00 |

7.  **Conclusions**

We conclude that from point of view of SNA, there are no significant correlations between relations of artists record sales on phonographic market with the view how people see music genres that artists belong to (Tables 2,3). Different groups of consumers, who built the market, do not buy in the same basket of records of similar artists, its extension to music genre is not allowed. The analysis of phonographic market based on the Social Network Analyzes revealed that there are two groups of customers. The first one buys music because of genres, and the second one buys record because of an artist. There is a different mechanism of consumer's decision making. There could be at least two possible explanations. The first come from classical theory of economy - marginal utility and product life-cycle management. Buying a new similar product has lower utility that previous one. It is well known that the main goal of record companies is to increase record sales. Thus, they often insist on releasing album on more convenient time, which maximizes income. Such a strategy could explain turbulence meaningful to the market (see Fig. 2). Unfortunately, the provided method to compare networks has some limitations. The biggest one is a lack of comparable time interval between quantitative algorithmic finding relations in 'popularity' and 'sales' network based on time interval (even those intervals differ), with 'similarities' based on human memory.

On the other hand, the peaks of record sales caused by new releases display analogy to turbulence in the fluid mechanics [10] rather than financial markets [19, 20]. This issue is worth investigating in future papers.